\def\spose#1{\hbox to 0pt{#1\hss}}
\def\approxlt{\mathrel{\spose{\lower 3pt\hbox{$\sim$}}
        \raise 2.0pt\hbox{$<$}}}
\def\approxgt{\mathrel{\spose{\lower 3pt\hbox{$\sim$}}
        \raise 2.0pt\hbox{$>$}}}

\def\multleft#1{\hbox to size{\vbox {\halign {\lft{##}\cr #1}}\hfill}\par}
\def\multright#1{\hbox to size{\vbox {\halign {\rt{##}\cr #1}}\hfill}\par}

\def\degmark{^\circ}
\def\today{\ifcase\month\or January\or February\or March\or April\or May\or
      June\or July\or August\or September\or October\or November\or December\fi
      \space\number\day, \number\year}
\def\s{\hbox{\phantom{5}}}

\def\boxit#1{\vbox{\hrule\hbox{\vrule\kern3pt\vbox{\kern3pt
          #1 \kern3pt}\kern3pt\vrule}\hrule}}

\def\cm{{\rm\thinspace cm}}

\def\erg{{\rm\thinspace erg}}
\def\eV{{\rm\thinspace eV}}

\def\keV{{\rm\thinspace keV}}
\def\km{{\rm\thinspace km}}

\def\Mpc{{\rm\thinspace Mpc}}

\def\s{{\rm\thinspace s}}

\def\ergpcmsqps{\hbox{$\erg\cm^{-2}\s^{-1}\,$}}

\def\ergps{\hbox{$\erg\s^{-1}\,$}}

\def\kmps{\hbox{$\km\s^{-1}\,$}}

\def\pcmsq{\hbox{$\cm^{-2}\,$}}

\def\ps{\hbox{$\s^{-1}\,$}}

\def\kmpspMpc{\hbox{$\kmps\Mpc^{-1}$}}

\documentstyle[psfig]{mn}

\begin{document}

\title[X-ray spectroscopy of 3C~111]{X-ray spectroscopy of the
broad line radio galaxy 3C~111}

\author[C.~S.~Reynolds et al.]
{{C.~S.~Reynolds$^{1,2}$, K.~Iwasawa$^2$, C.~S.~Crawford$^2$, A.~C.~Fabian$^2$}\\
\small{$^1$JILA, University of Colorado, Boulder, CO80309-0440, USA.}\\
\small{$^2$Institute of Astronomy, Madingley Road, Cambridge CB3 0HA.}}

\maketitle

\begin{abstract}
  We present an {\it ASCA} observation of the broad line radio galaxy
  3C~111.  The X-ray spectrum is well described by a model consisting of a
  photoelectrically-absorbed power-law form.  The inferred absorbing column
  density is significantly greater than expected on the basis of 21-cm
  measurements of Galactic {\sc H\,i}.  Whilst this may be due intrinsic
  absorption from a circumnuclear torus or highly warped accretion disk,
  inhomogeneities and molecular gas within the foreground giant molecular
  cloud may also be responsible for some of this excess absorption.  We
  also claim a marginal detection of a broad iron K$\alpha$ line which is
  well explained as being a fluorescent line originating from the central
  regions of a radiatively-efficient accretion disk.  This line appears
  weak in comparison to those found in (radio-quiet) Seyfert nuclei.  We
  briefly discuss the implications of this fact.
\end{abstract}


\section{Introduction}

Recent advances in X-ray spectroscopy, currently exemplified by the {\it
  ASCA} satellite (Tanaka, Holt \& Inoue 1994), have allowed new insights
into the astrophysics of the central engines of active galactic nuclei
(AGN).  Due to their proximity, and comparatively high brightness, the most
complete studies have focussed upon Seyfert galaxies.  Whereas the
underlying emission in Seyfert nuclei over the intrinsic {\it ASCA} band
(0.5--10\,keV) is found to be well approximated by a power-law form, most
Seyfert nuclei display additional spectral complexity.  Substantial
photoelectric absorption by cold and/or ionized circumnuclear material is
seen in half of all Seyfert 1 nuclei (Fabian et al. 1994; Reynolds 1997)
and all Seyfert 2 nuclei.  This circumnuclear material may be related to
the putative molecular torus, a warped accretion disk, and/or a disk-wind.
In addition, almost all Seyfert 1 nuclei are seen to possess a broad,
skewed iron K$\alpha$ emission line (Mushotzky et al. 1995; Tanaka et al.
1995; Nandra et al. 1997; Reynolds 1997).  This line is thought to be due
to fluorescence from the inner regions of the black hole accretion disk
that occurs when the disk is strongly irradiated by the X-ray emitting
plasma (Tanaka et al. 1995; Fabian et al. 1995).  These lines are typically
seen to have a full width at half maximum (FWHM) of $100\,000\kmps$ and
equivalent widths (EW) of $250\eV$.  A Compton backscattered continuum (the
so-called reflection continuum) accompanies these lines at approximately
the strength expected on the basis of the X-ray reprocessing model (George
\& Fabian 1991; Matt, Perola \& Piro 1991).

Detailed X-ray spectroscopic studies of radio-loud AGN are rather less
complete than that of Seyfert galaxies.  This is primarily due to the
comparative rarity of radio-loud objects (and, thus, the fact that we have
to observe them at greater distance).  However, this is a subject of some
importance: if the radio-loud/radio-quiet dichotomy reflects fundamental
differences in central engine structure, we would expect those differences
to be most directly revealed in the X-ray band.  The {\it ASCA} spectrum of
the powerful broad line radio galaxy (BLRG) 3C~109 reveals a fairly strong
and broad iron line with FWHM$\sim 120\,000\kmps$ and EW$\sim 300\eV$
(Allen et al. 1997).  This suggests a fundamental similarity between the
central engine of this radio galaxy and typical Seyfert nuclei, i.e. that
the central parts of the accretion flow are in the form of a
geometrically-thin, radiatively-efficient accretion disk.  However, 3C~109
might be the exception rather than the rule.  Zdziarski et al. (1995) and
Wo\'zniak et al. (1997) have shown that BLRGs tend to have slightly weaker
iron lines, and appreciably weaker reflection continua, as compared to
Seyfert nuclei.

In this paper, we report the results of an {\it ASCA} observation of 3C~111
($z=0.048$).  This is an X-ray bright BLRG which is classified as an FR-II
source with a double-lobe/single-jet morphology (Linfield \& Perley 1984).
The inner jet has been seen to display superluminal motion, with apparent
outflow velocities of $3.4c$ (Vermeulen \& Cohen 1994).  Assuming that the
pattern speed is intrinsically sub-luminal, the standard theory of
superluminal motion (e.g., see Blandford 1990) implies an upper limit on
the angle between the line of sight and the jet axis of
$\theta<32\degmark$.  

Section 2 reviews the previous X-ray observations of this source.  In
Section 3 we describe the basic {\it ASCA} data analysis and present the
results of our spectral fitting.  Section 4 discusses the astrophysical
implications of our findings.  Section 5 draws together our conclusions.
We assume that $H_0=50\kmpspMpc$ and $q_0=0.5$ throughout this work.

\section{A brief X-ray history of 3C~111}

\begin{table*}
  \begin{tabular}{lccccl}
Observatory \& & Observation & Photon & Absorbing column & 2-10\,keV Flux & Reference
\\
Instrument & Date & Index & ($10^{20}\pcmsq$) & ($10^{-11}\ergpcmsqps$) &
\\\hline
HEAO~1 & 1978-Sep-3 & $1.89^{+0.18}_{-0.17}$ & $194^{+101}_{-94}$ & 3.9
& Weaver et al. (1995) \\
Einstein SSS & 1979-Mar-2 & $1.47^{+0.47}_{-0.38}$ & $40^{+38}_{-26}$ &
unconstrained & HEASARC \\
Einstein MPC & 1979-Mar-2 & $1.56^{-0.50}_{+0.68}$ & $<252$ & 4.3 & HEASARC \\  
Einstein SSS/MPC & 1979-Mar-2 & $1.44^{+0.22}_{-0.20}$ & $36^{+29}_{-7}$ &
4.2 & Turner et al. (1991) \\
EXOSAT ME & 1983-Nov-14 & $1.35^{+0.31}_{-0.22}$ & $<66$ & 1.63 & HEASARC
\\
EXOSAT ME & 1984-Jan-27 & $2.71^{+1.50}_{-1.10}$ & $815^{+726}_{-473}$ &
0.97 & HEASARC \\  
EXOSAT LE/ME & 1984-Jan-27 & $1.4^{+0.9}_{-0.7}$ & $80^{+250}_{-50}$ & 1.8 &
Turner \& Pounds (1989) \\
Ginga LAC & 1989-Feb-4 & $1.77^{+0.04}_{-0.03}$ & $181^{+18}_{-13}$ &
3.1 & Nandra \& Pounds (1994) \\
ROSAT PSPC(B) & 1990-Jun-1 & $1.49^{+1.42}_{-1.20}$ & $67^{+39}_{-31}$ &
unconstrained & HEASARC \\  
ASCA SIS/GIS & 1996-Feb-13 & $1.72^{+0.05}_{-0.04}$ & $95^{+5}_{-4}$ &
3.5 & this work \\\hline
  \end{tabular}  
\caption{The X-ray history of 3C~111 including the current work.   Shown
  here are the results of fitting a simple absorbed power-law form to the
  relevant data.  Those datasets referenced as HEASARC have been retrieved
  from the {\sc HEASARC} database and fitted with an absorbed power-law
  model.  The errors are quoted at the 90 per cent level for one
  interesting parameters ($\Delta\chi^2=2.71$).}
\end{table*}

3C111 is a well known X-ray source and has been studied by every major
X-ray observatory since HEAO~1 A-2.  Since it appears as a point X-ray
source in every observation made to date, attention has focussed on
characterizing its spectral form.  To a good approximation, all X-ray
spectra of this object are consistent with a power-law spectrum modified by
the effects of neutral absorption.  Table~1 shows the results of fitting
such a model to various historical datasets for 3C~111.  Where possible,
the data have been retrieved from the High Energy Astrophysics Science
Archive Research Center ({\sc HEASARC}; located at the NASA-Goddard Space
Flight Center) and fitted using the appropriate response matrices and
background spectra.  For completeness, we also report the results of
published compilations even for those datasets that we have re-analyzed.

Comparing the detailed fits of these different instruments is wrought with
the dangers of cross-calibration uncertainties.  We do not attempt any such
comparison here and simply present Table~1 for completeness and
illustration.  However, we note one particularly interesting feature.  The
{\it EXOSAT} medium energy array (ME) observed 3C~111 on two occasions
separated by almost two and a half months.  It is evident from both the
fits in Table~1 and a visual examination of the two spectra that
significant spectral variability occurs between these two periods.  In
particular, the absorption column is seen to dramatically increase from
being less than $6.6\times 10^{21}\pcmsq$ to greater than $3.42\times
10^{22}\pcmsq$.  Since this significant change is seen in two spectra from
the same instrument, it appears to be a robust result.  We briefly discuss
the possible physical nature of this change in Section 4.

\section{Analysis of the {\it ASCA} data}

\subsection{Basic data reduction}

The {\it ASCA} observation of 3C~111 was performed on 1996 February 13/14
with a total duration of approximately one day.  Data were collected from
both the Solid-state Imaging Spectrometers (SIS) and Gas Imaging
Spectrometers (GIS).  The two SIS detectors were used in 1-CCD mode due to
telemetry constraints.  After applying standard data cleaning and selection
criteria (explicitly defined in Reynolds 1997), there was a total of
$32\,800\s$ of useful SIS data, and $38\,000\s$ of useful GIS data.  These
shall be referred to as `good' data.  Table~2 summarizes the basic
parameters of this observation.

\begin{table}
\begin{center}
\begin{tabular}{lc}
parameter & value \\\hline
obs. date & 13/14-Feb-1996 \\
good SIS time & 32\,800\,s \\
good GIS time & 38\,000\,s \\
SIS0 count rate & 0.702$\,{\rm ct}\,{\rm s}^{-1}$ \\
GIS2 count rate & 0.542$\,{\rm ct}\,{\rm s}^{-1}$ \\
0.5--2\,keV flux & $5.0\times 10^{-12}\ergpcmsqps$ \\
2--10\,keV flux & $3.5\times 10^{-11}\ergpcmsqps$ \\
0.5--2\,keV luminosity & $2.2\times 10^{44}\ergps$ \\
2--10\,keV luminosity & $3.9\times 10^{44}\ergps$ \\\hline
\end{tabular}
\end{center}
\caption{Basic parameters of the {\it ASCA} observation of 3C~111.   The
quoted fluxes are those observed (i.e. subject to the total line of sight
absorption), but the quoted luminosities are the intrinsic
(i.e. un-absorbed) values.  Spectral model-E of Table~3 was used to
determine the flux and luminosity.}
\end{table}

There is no evidence for spatial extent beyond that of the point spread
function in any of the images resulting from the four detectors.
Lightcurves and spectra were extracted using circular regions centred on
the centroid of the source counts.  We used an extraction radius of
3\,arcmins for the SIS data, and 4\,arcmins for the GIS data.  These
regions are sufficiently large to contain all but a small portion of the
source counts.  Background spectra were extracted from source free regions
of the same fields for each of the four detectors.  Background regions for
the SIS were taken to be rectangular regions along the edges of the source
chip, whereas annular regions were used for GIS background.  Following
standard practices, our spectra were rebinned such as to contain at least
20 photons per spectral bin.  This ensures that the photon number in each
spectral bin is approximately described by a normal distribution, a
necessary condition for $\chi^2$ analysis to be valid.  Table~2 reports the
resulting (background subtracted) count rates, fluxes and inferred
luminosities.

The light curves from each detector were examined for variability.  No
evidence for variability was found, i.e., $\chi^2$ analysis found that a
constant X-ray flux is completely consistent with the {\it ASCA} lightcurves.

\subsection{Spectral analysis}

In the absence of any spatial or temporal structure, we shall concentrate on
an analysis of the spectrum.  In order to avoid poorly calibrated regions
of the spectrum, we limited the SIS analysis to the 0.55--10\,keV band, and
GIS analysis to the 1--10\,keV band.  For convenience, the results of the
following analysis are summarized in Table~3.

Narrow beam 21-cm measurements suggest that the Galactic {\sc H\,i} column
density towards 3C~111 is $N_{\rm H}(Gal)=3.26\times 10^{21}\pcmsq$ (Elvis,
Lockman \& Wilkes 1989).  However, it must be noted that 3C~111 lies behind
a well-known Galactic molecular cloud.  Inhomogeneities within this cloud
on a scale smaller than that probed by the 21-cm measurements might lead to
differences between the real Galactic column density towards the X-ray
source in 3C~111, and that suggested by the {\sc H\,i} measurements.
Furthermore, molecular regions will contribute to the X-ray opacity but not
to the 21-cm flux.  We shall return to this issue in Section 4.

\begin{table*}
\begin{center}
\begin{tabular}{lllc}
model & model specifications & parameter values & $\chi^2$/dof \\\hline
A & NH(Gal)+PL & $N_{\rm H}(Gal)=3.26\times 10^{21}\pcmsq$ (fixed) & 5293/1725 \\
&           & $\Gamma=1.13\pm 0.01$ & \\\hline
B & NH(Gal)+NH+PL & $N_{\rm H}(Gal)=3.26\times 10^{21}\pcmsq$ (fixed) &
1653/1724 \\
&              & $N_{\rm H}=(6.2^{+0.3}_{-0.2})\times 10^{21}\pcmsq$ & \\
&              & $\Gamma=1.72^{+0.03}_{-0.02}$ & \\\hline
C & NH(Gal)+zNH+PL & $N_{\rm H}(Gal)=3.26\times 10^{21}\pcmsq$ (fixed) &
1659/1724 \\
&             & $N_{\rm H}=(7.0^{+0.2}_{-0.3})\times 10^{21}\pcmsq$ & \\
&              & $\Gamma=1.72\pm 0.02$ & \\\hline
D & NH(Gal)+WARM+PL & $N_{\rm H}(Gal)=3.26\times 10^{21}\pcmsq$ (fixed) &
1660/1723 \\
&             & $N_{\rm W}=(7.9^{+0.6}_{-1.2})\times 10^{21}\pcmsq$ & \\
&             & $\xi<7.3\ergps\cm$ & \\
&             & $\Gamma=1.65^{+0.03}_{-0.04}$ &\\\hline
E & NH(Gal)+NH+PL+nGAU & $N_{\rm H}(Gal)=3.26\times 10^{21}\pcmsq$ (fixed)
& 1649/1722 \\
&                      & $N_{\rm H}=(6.3^{+0.3}_{-0.2})\times
10^{21}\pcmsq$ & \\
&                      & $\Gamma=1.73\pm 0.02$ & \\
&                      & $E=6.4\pm 0.2\keV$ & \\
&                      & $W=20^{+25}_{-16}\eV$ & \\\hline
F & NH(Gal)+NH+PL+bGAU & $N_{\rm H}(Gal)=3.26\times 10^{21}\pcmsq$ (fixed)
& 1644/1721 \\
&                      & $N_{\rm H}=(6.4\pm 0.3)\times 10^{21}\pcmsq$ & \\
&                      & $\Gamma=1.75\pm 0.03$ & \\
&                      & $E=6.2^{+1.8}_{-0.9}\keV$ & \\
&                      & $\sigma=0.6^{+\infty}_{-0.5}\keV$ & \\
&                      & $W>20\eV$ & \\\hline
G & NH(Gal)+NH+PL+DISK & $N_{\rm H}(Gal)=3.26\times 10^{21}\pcmsq$ (fixed)
& 1643/1721 \\
&                      & $N_{\rm H}=(6.4\pm 0.3)\times 10^{21}\pcmsq$ & \\
&                      & $\Gamma=1.75\pm 0.03$ & \\
&                      & $r_{\rm in}=6r_{\rm g}$ (fixed) & \\
&                      & $r_{\rm out}=1000r_{\rm g}$ (fixed) & \\
&                      & $i<32\degmark$ & \\
&                      & $\beta<-2.2$ & \\
&                      & $W=100^{+95}_{-60}\eV$ & \\\hline
H & NH(Gal)+NH+PL+DISK & $N_{\rm H}(Gal)=3.26\times 10^{21}\pcmsq$ (fixed)
& 1643/1721 \\
&                      & $N_{\rm H}=(6.4\pm 0.3)\times 10^{21}\pcmsq$ & \\
&                      & $\Gamma=1.75\pm 0.03$ & \\
&                      & $r_{\rm in}<18r_{\rm g}$ & \\
&                      & $r_{\rm out}=1000r_{\rm g}$ (fixed) & \\
&                      & $i<32\degmark$ & \\
&                      & $\beta=-3$ (fixed) & \\
&                      & $W=110^{+60}_{-70}\eV$ & \\\hline
\end{tabular}
\end{center}
\caption{Spectral fitting results for 3C~111.   Model abbreviations follow:
  NH(Gal) = Galactic column density as inferred from {\sc H\,i}
  measurements, NH = additional (neutral) Galactic column density (column
  $N_{\rm H}$), zNH = additional (neutral) column density located at 3C~111
  (column $N_{\rm H}$), WARM = ionized (warm) absorber (column $N_{\rm W}$,
  ionization parameter $\xi$; see Section 3.2.1), PL = power-law emission
  (photon index $\Gamma$), nGAU = narrow Gaussian emission line (centroid
  energy $E$ and equivalent width $W$), bGAU = broad Gaussian emission line
  (centroid energy $E$, standard deviation $\sigma$ and equivalent width
  $W$).  DISK = diskline model (see Section 3.2.2 of main text for
  description).  All errors are quoted at the 90 per cent confidence level
  for one interesting parameter, $\Delta\chi^2=2.7$.}
\end{table*}

\begin{figure*}
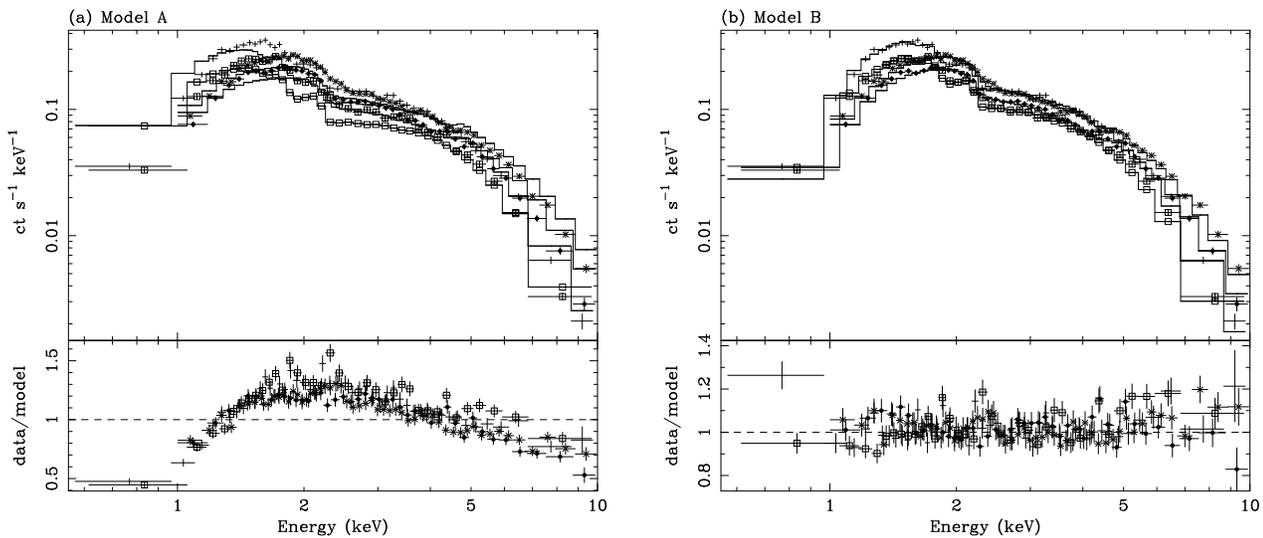

\hbox{
\psfig{figure=fig1a.ps,width=0.45\textwidth,angle=270}
\hspace{0.5cm}
\psfig{figure=fig1b.ps,width=0.45\textwidth,angle=270}
}
\caption{{\it ASCA} spectrum of 3C~111.  Panel (a) shows the data fitted to a
  power-law form modified only by Galactic absorption at the level
  suggested by {\sc Hi} 21-cm measurements (model A in Table~3).  Note the
  clear deficit of counts at soft energies indicating the need for extra
  absorption.  Panel (b) shows the data fitted with a model consisting of a
  power-law form modified by absorption due to a $z=0$ neutral absorber
  (model B in Table~3).  The data from all four instruments are shown: SIS0
  (plain crosses), SIS1 (open squares), GIS2 (filled circles) and GIS3
  (diagonal crosses).  For clarity of presentation, the spectral bins have
  been re-grouped by a factor of 20.}
\end{figure*}

\subsubsection{Basic spectral form and characterization of the X-ray absorption}

All spectral analysis reported in this paper is based upon joint fitting of
the SIS and GIS data.  The normalizations of the model for each of the four
instruments were left as free parameters to allow for the $\sim 10$ per
cent uncertainties in the overall flux calibration of the different
instruments.  However, we have noted a slight discrepancy between the SIS
and GIS data.  When taken alone, the SIS data present marginal evidence for
a spectral break at $4\keV$ with photon indices $\Gamma\approx 1.9$ below
the break and $\Gamma\approx 1.6$ above the break ($\Delta\chi^2=11$ for
two additional degrees of freedom as compared with the unbroken power-law).
However, this break is {\it not} required when the GIS data are also
considered.  Furthermore, the GIS data taken alone show no evidence for an
iron emission line whereas the SIS data do show marginal evidence.  Given
the marginal nature of these differences, and the remaining calibration
uncertainties present in the current analysis tools, we have chosen to take
the conservative approach and deal only with the joint SIS/GIS spectrum.

Initially, we fitted the joint spectrum with a power-law form modified only
by Galactic absorption at the level indicated by the 21-cm measurements.
As can be seen from Table~3, this fit is a dreadful description of the
data, yielding a goodness of fit parameter $\chi^2$/dof=5293/1725.
Inspection of the residuals clearly indicates the need for absorption in
excess of that indicated by the 21-cm measurements (see Fig.~1a).  Thus, we
consider additional column densities of neutral absorbing material placed
either in our Galaxy (i.e. $z=0$) or at 3C~111 (i.e. $z=0.048$).  Both of
these cases provided a statistically satisfactory description of the
overall spectrum, with required addition column densities of $N_{\rm
  H}=(6.2^{+0.3}_{-0.2})\times 10^{21}\pcmsq$ and $N_{\rm
  H}=(7.0^{+0.2}_{-0.3})\times 10^{21}\pcmsq$ for the $z=0$ and $z=0.048$
absorbers respectively.  Figure~1b shows the spectral fit for the case of
the $z=0$ absorber (although a very similar fit results for the $z=0.048$
absorber).

Inspection of Fig.~1 shows that the SIS0 data argue for extra soft emission
beyond that predicted by the model.  However, this conclusion is not
supported by data from SIS1.  Furthermore, examination of GIS2/GIS3 data
below 1\,keV (not shown on Fig.~1) also fails to provide any evidence for a
`soft excess'.  The level of this soft feature is much greater than the
SIS0 background level at those energies, and so we rule out the possibility
that poor background subtraction is responsible for this contradiction.  We
must conclude that this soft feature seen in SIS0 is not physical and
results from an extreme statistical fluctuation or some currently
unrecognized instrumental anomaly.

For completeness, we have also examined the possibility of absorption by
partially photoionized line-of-sight gas.  Such absorption is commonly seen
in the soft X-ray spectrum of Seyfert 1 galaxies where it is often referred
to as `warm' absorption.  We have constructed a photoionization model of an
incident power-law spectrum passing through a slab of gas with total column
density $N_{\rm W}$ and ionization parameter $\xi$.  This ionization
parameter is defined by
\begin{equation}
\xi=\frac{L_{\rm i}}{n_{\rm e}r^2},
\end{equation}
where $L_{\rm i}$ is the ionizing luminosity (i.e. the luminosity in
photons above $13.6\eV$), $n_{\rm e}$ is the electron number density of the
gas, and $r$ is the distance of the gas from the ionizing source.  For
detailed of this photoionization model, see Reynolds (1997) For a wide
range of parameter space, the ionization state of the gas, and hence its
X-ray opacity, depends only on $\xi$.  This is presented as model-D in
Table~3.

From table~3 it can be seen that the data do not require an ionized
absorber as is apparent from the fact that we can only impose an upper
limit on the ionization parameter $\xi<7.3\erg\cm\ps$.  Note that even this
upper limit is significantly below the typical values found in Seyfert
galaxies, $\xi\sim 20\erg\cm\ps$.

\subsubsection{The iron line}

The K$\alpha$ iron emission line is of some importance since it probes
structures deep within the central engine.  Here, we examine the 3C~111
data for evidence of such a line.

Throughout this section, we take the simple absorbed power-law model
(model~B in Table~3) as our base model.  To this base model, we added a
narrow Gaussian feature at $E=6.4\keV$ (in the rest frame of 3C~111) and
performed a $\chi^2$ minimization (leaving the energy $E$ and normalization
of the line as free parameters, but constraining the line to be narrow).
The results are given in Table~3 (model E).  The improvement in the
goodness of fit, $\Delta\chi^2=4.3$ is {\it not} significant at the 90 per
cent level according to the F-test for two extra degrees of freedom.
However, we obtain a stronger result if we add a broad Gaussian line to
model~B (reported as model F in Table~3): the improvement in the goodness
of fit parameter, $\Delta\chi^2=9$, is significant at the 90 per cent
level, but not at the 95 per cent level.  Thus, we claim a marginal
detection of a broad iron line.  However, due to the marginality of the
detection, the best fitting parameters of the broad Gaussian are very
poorly constrained, as can be seen by inspection of Table~3.

To make further progress, we assume that this line arises by the
fluorescence of `cold'\footnote{By `cold', we mean in the ionization range
Fe\,{\sc i}--Fe\,{\sc xvii}, giving a rest frame line energy of $6.4\keV$.}
iron in an accretion disk around a Schwarzschild black hole.  The expected
line profiles for this case have been calculated by Fabian et al. (1989)
and are incorporated into the spectral fitting package {\sc xspec} as the
`diskline' model.  

We considered two cases.  In the first case (model~G), the inner radius of
the line emitting region was fixed to correspond to the radius of marginal
stability $r_{\rm in}=6r_{\rm g}$, where $r_{\rm g}$ is the gravitational
radius of the black hole.  The outer radius of the line emitting region was
fixed at $r_{\rm out}=1000r_{\rm g}$.  We can justify fixing this parameter
since the line profile is insensitive to $r_{\rm out}$ provided that
$r_{\rm out}\gg r_{\rm in}$ and the line emissivity declines with radius in
the disk faster than $r^{-2}$.  We make the canonical simplification that
the line emissivity depends on the radius in a power-law sense,
$\epsilon\propto r^{\beta}$, where $\beta$ is left as a free parameter in
the fit.  The inclination of the disk, $i$, and normalization of the line
were also left as free parameters in the fit.  Our second case (model~H) is
similar except that we fix the emissivity index to be $\beta=-3$ and allow
the inner radius of the line emitting region to be a free parameter.  This
line emissivity law ($\beta=-3$) is the asymptotic behaviour for large $r$
in the case where the primary X-ray source is either a point X-ray source
on the symmetry axis of the disk, or is a thin corona with a local power
that tracks the viscous dissipation in the underlying disk (Page \& Thorne
1974).

The results of spectral fitting with models G and H are shown in Table~3.
Several of these results are noteworthy.  First, both models imply a disk
inclination of $i<32\degmark$, completely consistent with the limits on the
angle between the line of sight and the jet axis imposed by observations of
the superluminal motion (see Introduction).  Secondly, interesting
constraints can be placed on the equivalent width of the iron line,
$W=40-195\eV$ (90 per cent bounds).  This contrasts with the broad Gaussian
parameterization, with which only a lower limit on $W$ could be obtained.
The reason for this is easily seen.  The broad Gaussian fit is subject to a
statistical instability in which both the width and equivalent width of the
Gaussian grow arbitrary large.  In other words, the Gaussian component can
mimic the high-energy region of the continuum making meaningful constraints
on the EW of the line impossible.  However, the disk-line model imposes a
physical limit on the maximum width of the iron line, thereby allowing the
equivalent width to be usefully constrained.  Thirdly, model~H shows that
the line emitting region must extend at least down to within $18r_{\rm g}$
of the black hole.

\section{Discussion}

These {\it ASCA} data confirm the need for a substantial amount of
absorption beyond that expected on the basis of 21-cm Galactic {\sc H\,i}
measurements.  However, these data cannot distinguish between an absorber
which is intrinsic to the 3C~111 system (i.e., at $z=0.048$) or one that is
more local to us.  The small difference in the goodness of fit between
model B (i.e. the local absorber) and model C (i.e., the absorber at
$z=0.048$) may be influenced by the small soft excess seen in the SIS0
data.  Since we presume this soft excess to be spurious (Section 3.2.1), we
must treat model B and model C as providing similarly good fits to these
data.

Given this freedom in the spectral data, there are at least three
possibilities for the nature of this absorber.  First, the excess absorbing
material may be the neutral component of the ISM of the 3C~111 host galaxy.
Given the large column of material required (almost $10^{22}\pcmsq$), this
would suggest that the host galaxy was a disk galaxy viewed at high
inclination.  This is inconsistent with the apparent cD nature of the host
galaxy (Owen \& Laing 1989).  Thus, we argue against this possibility.

Secondly, the excess absorbing material maybe circumnuclear material
associated directly with the AGN (e.g., the putative molecular torus of AGN
unification schemes or the outer parts of a warped accretion disk.)
Observations of the iron line and, independently, superluminal motion imply
a central engine inclination of $i<32\degmark$ (i.e., we are viewing the
central engine somewhat face-on).  Within the ``circumnuclear absorber''
picture, we would be led to the conclusion that a relatively large quantity
of neutral material must exist at high latitudes within the system.  This
would suggest a molecular torus with a small opening angle or a highly
warped accretion disk.

In most cases, X-ray absorption towards an AGN in excess of that suggested
by Galactic {\sc H\,i} measurements is taken to be firm evidence for
intrinsic absorption associated with either the AGN or its host galaxy.
Whilst the absorbing material seen towards 3C~111 may indeed be
circumnuclear in nature, we might question this inference given the fact
that 3C~111 lies behind a well known giant molecular cloud (containing
Taurus~B and the Taurus-Perseus complexes) estimated to lie at a distance
of 350\,pc (Ungerechts \& Thaddeus 1987).  There are two ways in which the
presence of this cloud complex can cause deviations between the actual
Galactic absorption along the line of sight and that inferred from 21-cm
measurements: small scale inhomogeneities and the presence of molecular
hydrogen.   We note that Turner et al. (1995) made a similar suggestion for
the case of NRAO~140.

Marscher, Moore \& Bania (1993) and Moore \& Marscher (1995) have used
3C~111 as a background radio continuum source to probe details of the
4.83\,GHz H$_2$CO absorption line arising from material in this molecular
cloud.  On the basis of temporal variability in the strength and profile of
the absorption line, they argue that the cloud complex contains
inhomogeneities on the sub-parsec, or even AU, scales\footnote{We note that
  Thoraval, Boisse \& Stark (1996) dispute the existence of AU scale
  inhomogeneities in this cloud based on the apparent lack of reddening
  variability in a sample of stars in this field.}.  These authors estimate
that between 30--100 clumps lie along the line-of-sight to 3C~111 and that
Poisson fluctuations in this number give rise to the observed molecular
line changes.  The 21-cm studies of Elvis, Lockman \& Wilkes (1989) use a
$21\times 21$\,arcmin$^2$ beam (Lockman, Jahoda \& McCammon 1986) --
measurements of the neutral hydrogen column densities are necessarily
averaged over this beam and hence insensitive to the small scale
inhomogeneities indicated by the molecular absorption line studies.  Thus,
these inhomogeneities could give rise to differences between the 21-cm
column density and observed absorbing column.  However, given that 3C~111
appears {\it not} to be blocked by one (or a small number) especially
dense clump, it is difficult to envisage the factor of 3 discrepancy
between the 21-cm measurements and X-ray absorption as arising purely from
these inhomogeneities.

Clearly, we expect there to be molecular hydrogen along our line of sight
to 3C~111 associated with the molecular cloud.  This gas, and the
metals/dust associated with it, will act as a source of X-ray opacity which
is indistinguishable from atomic gas at the spectral resolution of {\it
  ASCA}.  Since it does not contribute to the 21-cm emission, this
molecular gas is another possible cause for the discrepancy between the
21-cm measurements and the X-ray absorbing column.  If 60--70 per cent of
the gas along the line of sight to 3C~111 is in molecular form, the
discrepancy between the 21-cm measurements and the X-ray absorbing column
is resolved.   Furthermore, the clear temporal change in absorption between
the two {\it EXOSAT} ME spectra discussed in Section 2 might be due to
a particularly large inhomogeneity/knot in this molecular gas that drifts
across the line of sight to the core of 3C~111.

We also claim the marginal detection of a broad iron line.  Although the
constraints are poor, the data are consistent with a cold iron line
originating from the central regions of an accretion disk around a
Schwarzschild black hole.  The disk inclination is constrained to be
$i<32\degmark$.  The equivalent width, $W=40-195\eV$, is rather weak as
compared to Seyfert nuclei (Nandra et al. 1997; Reynolds 1997).  This is
consistent with the results of Wo\'zniak et al. (1997) who examine {\it
OSSE}/{\it Ginga} data and find an iron line EW of $\sim 70\eV$ and a weak
reflection continuum.  This authors find that the inferred solid angle
subtended by the `reflector' at the X-ray source is $\Omega/2\pi\sim 0.3$.
As mentioned in the Introduction, this is found to be a general trend
differentiating radio-quiet Seyfert nuclei from BLRGs.

These iron line studies suggest both fundamental similarities and
differences between the central engines of radio-quiet Seyfert nuclei and
BLRGs.  The fact that BLRGs display resolved iron lines at all is evidence
for the existence of a geometrically-thin, radiatively efficient accretion
disk extending down to $r\approxlt 20r_{\rm g}$, as is the case for Seyfert
nuclei (note that in the Seyfert case, iron line studies seem to suggest
that the disk remains geometrically-thin and radiatively-efficient
essentially all of the way down to the black hole).  The comparative
weakness of the lines/reflection-continua in BLRG could arise for several
reasons.  First, there might be a component of the X-ray emission that is
associated with the radio jet and is beamed along the jet axis thereby
diluting the Seyfert-like reflection features that are intrinsically
present.  Secondly, the inner accretion disk (i.e. the inner $20r_{\rm g}$
or so) might not be in a physical state capable of producing the X-ray
reflection features.  For example, the disk might undergo a transition into
an optically-thin advection dominated disk (ADD) which would be totally
ionized and hence incapable of producing X-ray reflection signatures.  If
most of the primary X-ray flux still originated from the central most
regions, the solid angle covered by material capable of producing
reflection features could be rather small.  Such models (with a
thin-disk/ADD transition) have previously been discussed in the context of
Galactic Black Hole Candidates (e.g., see Esin, McClintock
\& Narayan 1997 and references therein).

\section{Conclusions}

Our {\it ASCA} data for 3C~111 clearly reveal the presence of absorption in
excess to that suggested by 21-cm measurements of the Galactic {\sc H\,i}
column.  Whilst this may be evidence for absorption intrinsic to the 3C~111
system, we suggest that inhomogeneities and molecular
material in the foreground (Galactic) giant molecular cloud may also be
responsible for this excess absorption.  Molecular absorption line
studies of this cloud lend strength to this possibility.

We also claim the marginal detection of a broad iron K$\alpha$ line.  The
detection of a broad line at all suggests the existence of a
geometrically-thin, radiatively-efficient accretion disk within $r\approxlt
20GM/c^2$ of the black hole.  However, the weakness of this line (and the
associated backscattered continuum) as compared to typical Seyfert galaxies
suggests that either X-ray beaming is important, or that the innermost
regions of the disk are in a state incapable of producing the X-ray
reflection signatures.  One possibility is that the disk undergoes a
transition to an ADD state in its innermost regions.

\section*{Acknowledgments}

We acknowledge support from PPARC (CSR, CSC, KI), the Royal Society (ACF),
the National Science Foundation under grant AST9529175 (CSR), and NASA
under grant NASA-NAG5-6337 (CSR).


\begin{thebibliography}{}
\bibitem{} Allen S.~W., Fabian A.~C., Idesawa E., Inoue H., Kii T., Otani
C., 1997, MNRAS, 286, 765 
\bibitem{} Blandford R.~D., 1990, in Active Galactic Nuclei, ed
T.J.-L.Courvoisier \& M.Mayor (Saas-Fee Advanced Course 20)
(Berlin:Springer), p.161
\bibitem{} Elvis M., Lockman F.~J., Wilkes B.~J., 1989, AJ, 97, 777
\bibitem{} Esin A.~A., McClintock J.~E., Narayan R., 1997, ApJ, submitted
\bibitem{} Fabian A.~C., Rees M.~J., Stellar L., White N.~E., 1989, MNRAS, 238, 729
\bibitem{} Fabian A.~C. et al. 1994, PASJ, 46, L59
\bibitem{} Fabian A.~C. et al. 1995, MNRAS, 277, L11
\bibitem{} George I.~M., Fabian A.~C., 1991, MNRAS, 249, 352
\bibitem{} Linfield R., Perley R., 1984, ApJ, 279, 60
\bibitem{} Lockman F.~J., Jahoda K., McCammon D., 1986, ApJ, 302, 432
\bibitem{} Marscher A.~P., Moore E.~M., Bania T.~M., 1993, ApJ, 419, L101
\bibitem{} Matt G., Perola G.~C., Piro L., 1991, A\&A, 247, 25
\bibitem{} Moore E.~M., Marscher A.~P., 1995, ApJ, 452, 671
\bibitem{} Mushotzky R.~F., Fabian A.~C., Iwasawa K., Kunieda H., Matsuoka
M., Nandra K., Tanaka Y., 1995, MNRAS, 272, L9
\bibitem{} Nandra K., Pounds K.~A., 1994, MNRAS, 268, 405
\bibitem{} Nandra K., George I.~M., Mushotzky R.~F., Turner T.~J., Yaqoob
T., 1997, ApJ, 477, 602
\bibitem{} Owen F.~N., Laing R.~A., 1989, MNRAS, 238, 357
\bibitem{} Page D.~N., Thorne K.~S., 1974, ApJ, 499, 191
\bibitem{} Reynolds C.~S., 1997, MNRAS, 286, 513
\bibitem{} Tanaka Y., Inoue H., Holt S.~S., 1994, PASJ, 46, L37
\bibitem{} Tanaka Y. et al.,  1995, Nat, 375, 659
\bibitem{} Thoraval S., Boisse P., Stark R., 1996, A\&A, 312, 973
\bibitem{} Turner T.~J., Pounds K.~A., 1989, MNRAS, 240, 833
\bibitem{} Turner T.~J., Weaver K.~A., Mushotzky R.~F., Holt S.~S.,
  Madejski G.~M., 1991, ApJ, 381, 85 
\bibitem{} Turner T.~J., George I.~M., Madejski G.~M., Kitamoto S., Suzuki
  T., 1995, ApJ, 445, 660
\bibitem{} Ungerechts H., Thaddeus P., 1987, ApJS, 63, 645
\bibitem{} Vermeulen R.~C., Cohen M.~H., 1994, ApJ, 430, 467
\bibitem{} Weaver K.~A., Arnaud K.~A., Mushotzky R.~F., 1995, ApJ, 447, 121
\bibitem{} Wo\'zniak P.~R., Zdziarski A.~A., Smith D., Madejski G.~M.,
Johnson W.~N., 1997, MNRAS, submitted
\bibitem{} Zdziarski A.~A., Johnson W.~N., Done C., Smith D., McNaron-Brown
K., 1995, ApJ, 438, L63
\end{thebibliography}
\end{document}